\documentclass[12pt,a4paper]{article}
\usepackage{graphicx}
\usepackage{amsmath,amssymb}
\usepackage{amsfonts}
\begin{document}
\begin{center}

{\bf NUCLEAR MATRIX ELEMENTS FOR DOUBLE BETA DECAY}

 \medskip
Vadim Rodin

 \medskip
{\it
Institut f\"{u}r Theoretische Physik der Universit\"{a}t T\"{u}bingen, D-72076 T\"{u}bingen, Germany
}
\end{center}

\vspace*{-1cm}

\begin{abstract}
The present status of calculations of the nuclear matrix elements for neutrinoless double beta decay is reviewed.
A proposal which allows in principle to measure the neutrinoless double beta decay Fermi matrix 
element is briefly described.
\end{abstract}

\bigskip

\section{Introduction}

Neutrino oscillation experiments have proven that neutrinos are massive particles (see, e.g., Ref.~\cite{McK04}). 
However, the absolute scale of the neutrino masses cannot in principle be deduced from the observed oscillations.
To determine the absolute neutrino masses down to the level of tens of meV,
study of the neutrinoless double beta ($0\nu\beta\beta$) decay~~$^A_Z {\mathrm{El}}_N \longrightarrow \ _{Z+2}^{\phantom{+2} A} {\mathrm{El}}_{N-2} + 2e^-$ 
becomes indispensable.
Furthermore, this process, which violates the total lepton number by two units, is an
{\em experimentum crucis} to reveal the Majorana nature of neutrinos, i.e. if neutrino is identical with antineutrino (see, e.g., Ref.~\cite{vogelbook,fae98,Suh98,AEE07}).

Determination of the effective Majorana mass (or relevant GUT and SUSY parameters depending on 
what mechanism of the $0\nu\beta\beta$ \ decay dominates) 
from experimental data on the $0\nu\beta\beta$-decay lifetimes can be only as good as 
the knowledge of the nuclear matrix elements $M^{0\nu}$ on which the $0\nu\beta\beta$ decay rates depend. 
Thus, a better understanding of the nuclear structure effects important for describing the matrix
elements is needed to interpret the future data accurately. 
It is crucial in this connection to develop theoretical methods capable of reliably evaluating 
the nuclear matrix elements, and to realistically assess their uncertainties. 

In general, barring contributions different from light Majorana neutrino exchange, 
the inverse $0\nu\beta\beta$ lifetime in a given nucleus is the product of three factors,
\begin{equation}
\label{3fact}
\left(T^{0\nu}_{1/2}\right)^{-1}=G^{0\nu}\,\left|M^{0\nu}\right|^2\,m_{\beta\beta}^2\ ,
\end{equation}
where $G^{0\nu}$ is a calculable phase space factor, $M^{0\nu}$ is the $0\nu\beta\beta$ nuclear
matrix element, and $m_{\beta\beta}$ is the so-called
``effective Majorana neutrino mass'' which, in standard notation~\cite{PDGr}, reads
$m_{\beta\beta}=\left|\sum_{1=1}^3 m_i\,U^2_{ei}\right|\ ,$
with $m_i$ and $U_{ei}$ being the neutrino masses and the $\nu_e$ mixing matrix elements,
respectively.

Two-neutrino double beta ($2\nu\beta\beta$) decay, 
$^A_Z {\mathrm{El}}_N \longrightarrow \ _{Z+2}^{\phantom{+2} A} {\mathrm{El}}_{N-2} + 2e^-+2\overline\nu_e$
is a second-order weak process which probes the same mother and daughter
nuclei as $0\nu\beta\beta$ decay. It is allowed within the standard model and has been observed in several nuclei~(see, e.g.,~\cite{barab}). The decay provides a particularly important benchmark. As
it was extensively demonstrated in~\cite{Rod03a}, the spread of QRPA 
calculation results for $M^{0\nu}$ can be significantly reduced by constraining the nuclear model
with the corresponding experimental $2\nu\beta\beta$ decay lifetimes.

The calculation of the matrix element $M^{0\nu}$ for a candidate $0\nu\beta\beta$
nucleus is notoriously difficult. It requires the detailed description of a second-order 
weak decay from a double-even mother nucleus $(Z,\,A)$ to a double-even daughter nucleus $(Z+2,\,A)$ via virtual states (with any multipolarity $J^\pi$) of the intermediate nucleus $(Z+1,\,A)$.

Two basic methods are used in the evaluation of $M^{0\nu}$, the quasiparticle random phase approximation (QRPA),
with its various modifications~\cite{Rod03a,anatomy,src09,Rod08,Kort07} and the nuclear
shell model (NSM)~\cite{cau08} (very recently $M^{0\nu}$ have also been calculated within the IBM~\cite{Bar09}). 
The NSM aims at complete describing the nuclear wave functions 
by taking into account nucleon configurations of all possible complexity. By diagonalizing the nuclear Hamiltonian 
within the model space the energies of ground and excited states of nucleus as well as the corresponding wave functions can be calculated. However, application of the NSM to description of the medium and heavy nuclei beyond the $pf$-shell immediately faces factorial growth of the dimension of the model space . Therefore, severe truncation of the size of the single-particle basis is usually made in the medium and heavy nuclei. 

In contrast to the NSM, within the QRPA and the renormalized QRPA (RQRPA)
one can include essentially unlimited set of single-particle states, 
but only a limited subset of configurations (iterations of the particle-hole, 
respectively two-quasiparticle configurations) is taken into account.  
On the other hand, within the QRPA there is no obvious procedure  
that determines how many single particle states one should include. 
Hence, various authors choose this crucial number {\it ad hoc}, basically 
for reasons of convenience.  


\section{Current status of calculations of $M^{0\nu}$}

The $0\nu\beta\beta$ nuclear matrix element is a sum of Fermi (F), Gamow-Teller (GT) transitions and (a small) tensor (T) contributions (see, e.g.,~\cite{Rod03a} for detailed representations of the contributions),
\begin{equation}
M^{0\nu} = M^{0\nu}_\mathrm{GT}+M^{0\nu}_\mathrm{T}-\displaystyle\frac{M_\mathrm{F}^{0\nu}}{g^2_A}\ ,
\end{equation}
and nuclear models are needed to estimate the different components
$M^{0\nu}_\mathrm{X}$ ($\mathrm{X}=\mathrm{F}$, GT, T). In the above expression,
 $g_A$ is the effective axial coupling in nuclear matter, not necessarily equal 
to its ``bare'' free-nucleon value $g_A\simeq1.25$. A direct comparison of $M^{0\nu}$ 
calculated for different $g_A$ can be done 
in terms of the matrix element ${M'}^{0\nu}=\left(\frac{g_A}{1.25}\right)^2 M^{0\nu}$.

\begin{figure}[tb]
\includegraphics[scale=0.6]{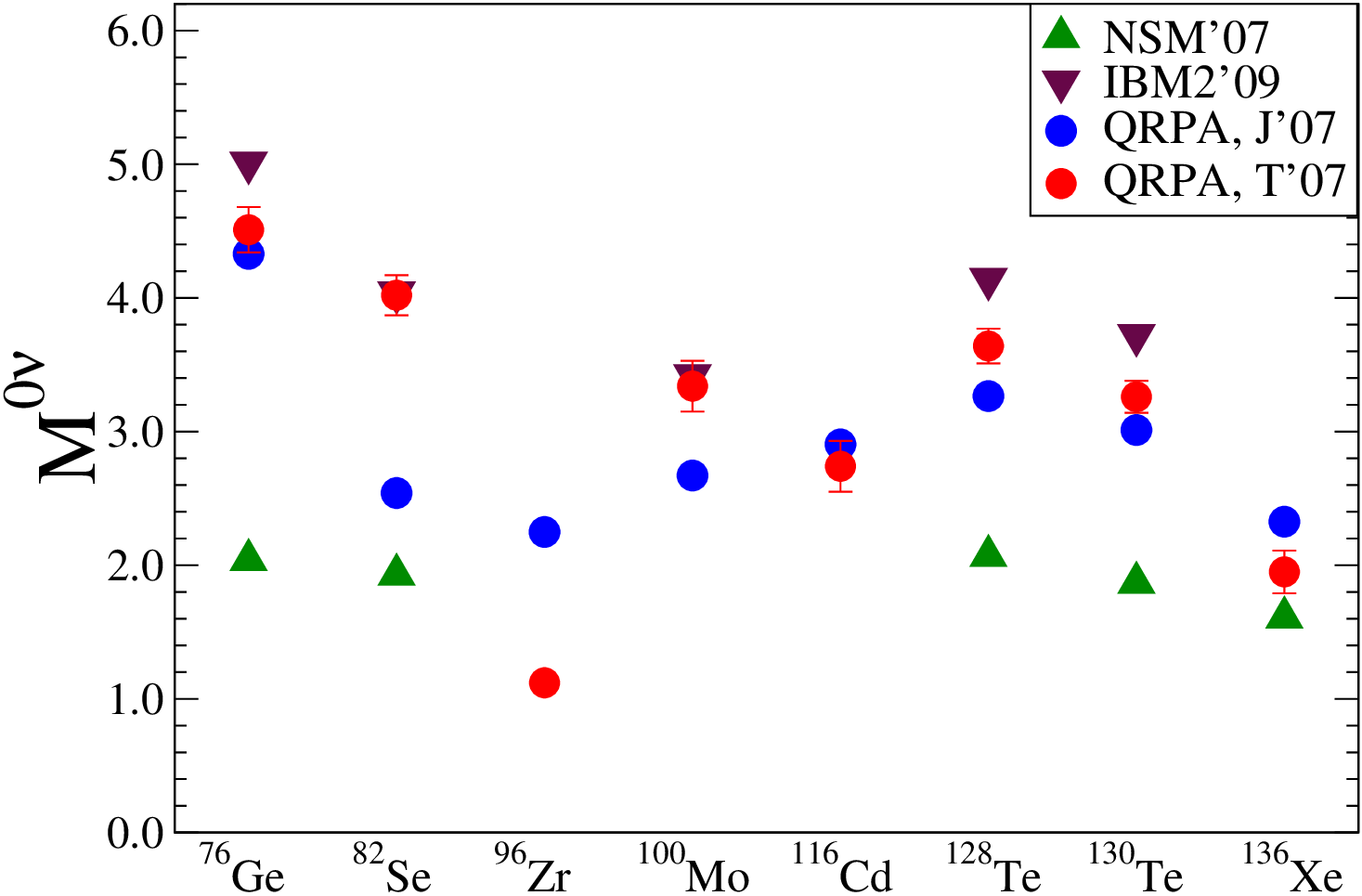}
\caption{Current status of calculations of $M^{0\nu}$ for the light neutrino 
exchange mechanism (with the Jastrow-like s.r.c. and $g_A=1.25$)
within different nuclear structure models (QRPA:  \cite{Rod03a} (T'07) and \cite{Kort07}  (J'07); NSM~\cite{cau08};
IBM2 \cite{Bar09}).}
\label{status}
\end{figure}

There has been great progress in the calculations of $M^{0\nu}$ over the last five years.
A comparison of the results 
by different groups (with the Jastrow-like short-range correlations (s.r.c.) and with the unquenched value $g_A=1.25$) is represented in Fig.~\ref{status}.
One can see in the figure that the matrix elements $M^{0\nu}$ of different groups calculated within 
the QRPA seem to converge. At the same time, the $M^{0\nu}$ of the NSM are systematically and substantially smaller (up to a factor of 2 for lighter nuclei like $^{76}$Ge) than the corresponding QRPA ones. There is now an active discussion in literature on what could be the reason of such a discrepancy, 
a too small single-particle model space of the NSM or a neglect of complex nuclear configurations within the QRPA.
The recent results obtained within the IBM method~\cite{Bar09} agree surprisingly well with the QRPA ones of Ref.~\cite{Rod03a}.

As already mentioned, all models employ truncations: the NSM severely truncates the s.p. model space whereas the QRPA does so with respect to the configuration space. The question relevant for calculation of the $0\nu\beta\beta$ nuclear matrix elements is which truncation induces the smallest error in $M^{0\nu}$.

In Ref.~\cite{cau08} the difference between the NSM and the QRPA results is attributed to a neglect of 
a subset of ground state correlation in the latter. 
The claim as it appears in  \cite{cau08} is ``the QRPA can be said to be a ''low seniority approximation",
roughly equivalent to the $s\le4$ ISM truncations, that overestimate the NME's ...". 
However, it is obviously incorrect that all the components with the seniority $s>4$ are not included within the QRPA since an analytic expression for the QRPA ground state as a coherent state built on top of the BCS vacuum can explicitly be derived (see, e.g.,~\cite{RingSchuck80}) This representation shows that all the configurations with the seniority $4N$ (where $N=0,1,2,\dots$) are taken into consideration when the QRPA ground state is calculated. 
Of course, those represent only a part of the entire ground state correlations 
but the part which is the most relevant for calculating the transition amplitudes. 

From our point of view it is namely too small single-particle model space used in the NSM calculations that is
responsible for the suppression of the calculated NSM $M^{0\nu}$. Only very limited number of negative parity configurations can be constructed within the $0\hbar\omega$ model space. Therefore, contribution from 
many $0\nu\beta\beta$-transitions (dipole, spin-dipole etc.)  via negative parity intermediate states 
are missing in the NSM description. Such transitions contribute a lot to the $M^{0\nu}$ as demonstrated by the QRPA results.
Thus, it is natural to expect the NSM matrix elements $M^{0\nu}$ to come out small because some important transitions contributing to the matrix element simply cannot be described in such a small basis. 

\section{QRPA analysis of uncertainties in $M^{0\nu}$}
At present, the most elaborate analysis of uncertainties in the $0\nu\beta\beta$ decay nuclear matrix elements $M^{0\nu}$ calculated within the QRPA and the RQRPA has been performed in Refs.~\cite{Rod03a,anatomy,src09}. 
Single-particle model spaces comprising $N=$2, 3 and 5 major oscillator shells were used in the (R)QRPA calculations along with different representations of the short range correlations.
The experimental $2\nu\beta\beta$ decay rates 
were used to adjust the most relevant parameter, the strength $g_{pp}$ of the particle-particle
interaction, and thus to ``calibrate'' the QRPA estimates of $M^{0\nu}$. The major observation of 
Refs.~\cite{Rod03a,anatomy,src09} is that 
such a procedure makes the calculated $M^{0\nu}$ essentially independent of the size of the single-particle basis of the QRPA. Furthermore, the matrix elements have been demonstrated to also become rather stable with respect to the possible quenching of the axial vector coupling constant $g_A$. 

Despite the fact that the $2\nu\beta\beta$ decay process probes only a subset of the  intermediate
states relevant for $0\nu\beta\beta$ decay (i.e., only those with $J^\pi=1^+$, via GT transitions), it is just
the $1^+$ contribution to the total $0\nu\beta\beta$ matrix element that reveals a pronounced 
sensitivity to $g_{pp}$, in contrast to the other multipole contributions~\cite{Rod03a}. 
This observation justifies the aforementioned fitting procedure employed in Ref.~\cite{Rod03a}.

\begin{figure}[tb]
\includegraphics[scale=0.6]{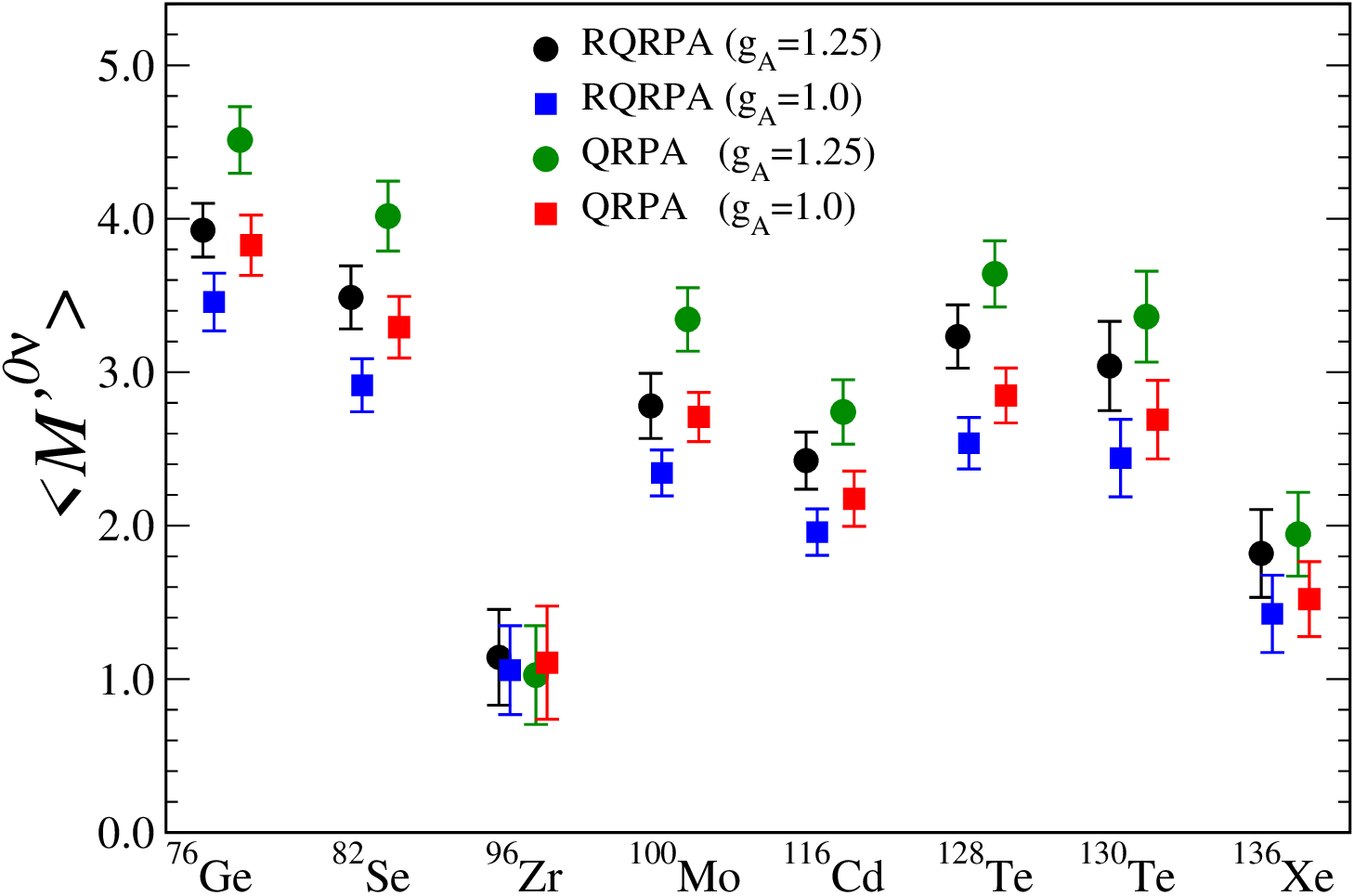}
\caption{Average nuclear matrix elements $\langle {M'}^{0\nu} \rangle $ 
and their variance (including the error coming from the experimental uncertainty in $M^{2\nu}$) 
for both the QRPA and the RQRPA~\cite{Rod03a}.}
\label{pme}
\end{figure}

The matrix elements ${M'}^{0\nu}$ calculated for the three 
single-particle bases and a fixed $g_A$ are relatively close to each other. 
For each nucleus the corresponding average  $\langle {M'}^{0\nu} \rangle$ matrix elements  
(averaged over the three choices of the single-particle space) was evaluated in Ref.~\cite{Rod03a}, 
as well as its variance $\sigma$. 
The final (R)QRPA results obtained by using unquenched and quenched values of $g_A$
($g_A=1.25$ and $g_A=1.0$, respectively) 
are presented in graphical form in  Fig.~\ref{pme}. The full uncertainties of the calculated $ {M'}^{0\nu}$
includes also the ones induced by the experimental uncertainties in $M^{2\nu}$.
One can see that not only is the variance substantially less than the average value, but the results of QRPA
are quite close to the RQRPA values.

\begin{figure}[tb]
\includegraphics[scale=0.6]{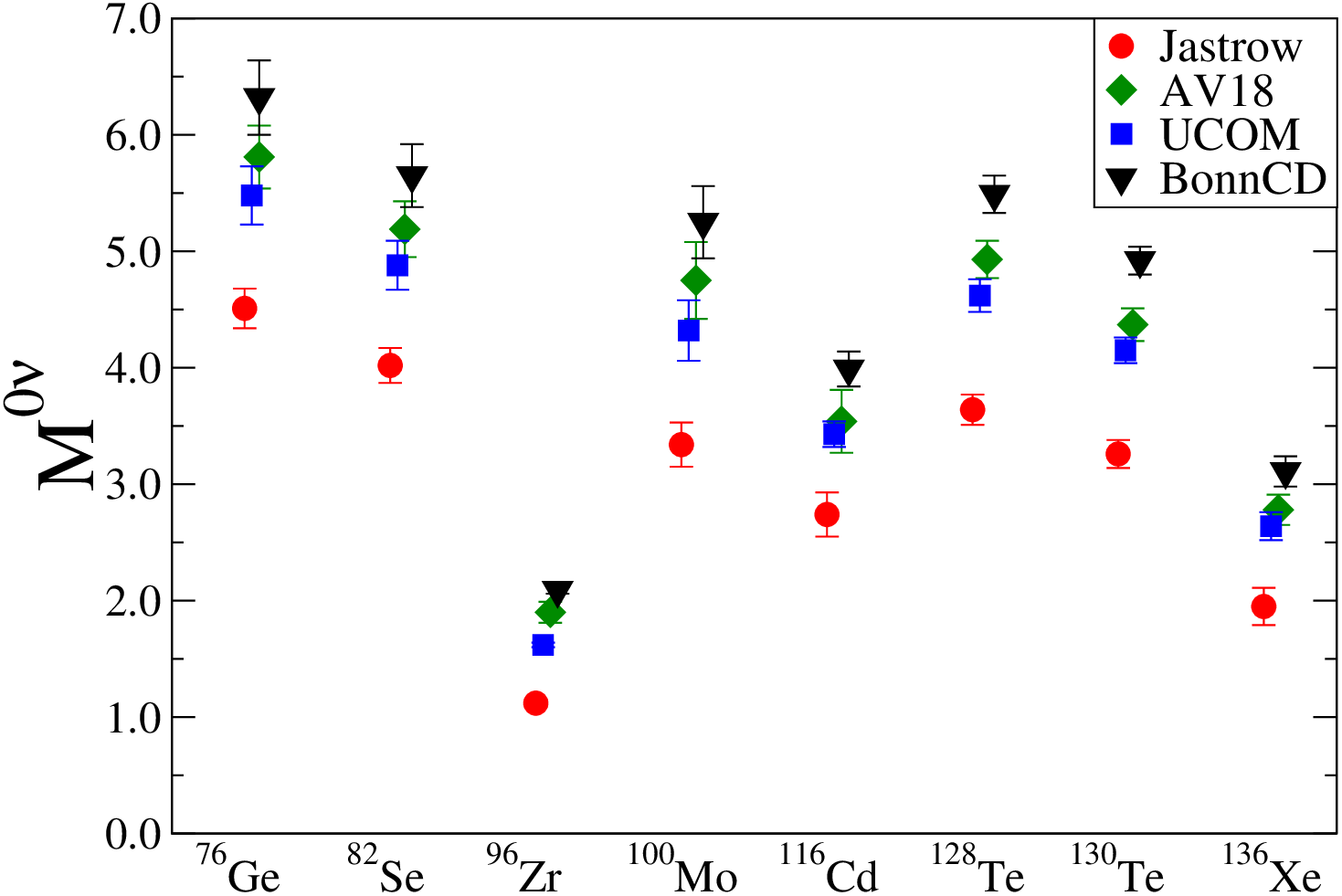}
\caption{Effect of different choices of the s.r.c. on $M^{0\nu}$ calculated within the QRPA~\cite{Rod03a,anatomy,src09}
($g_A=1.25$).}
\label{src}
\end{figure}

A systematic analysis of the effect of different choices of the s.r.c. on $M^{0\nu}$ calculated within the QRPA and the RQRPA has been performed in Refs.~\cite{Rod03a,anatomy,src09}. The QRPA results obtained with the Jastrow-like and the unitary correlation operator method (UCOM) treatment of the s.r.c. are represented in Fig.~\ref{src}. 
Also shown in the figure are the results of the first self-consistent calculation~\cite{src09} which uses the 
residual nuclear interactions as well as the s.r.c. derived from the same modern realistic nucleon-nucleon potentials, namely from charge-dependent Bonn potential (CD-Bonn) and the Argonne V18 potential. 
Larger matrix elements $M^{0\nu}$ are obtained as compared with the traditional approach of using the Jastrow-like treatment of the s.r.c..

\section{Can one measure $M^{0\nu}$?}
Given the uncertainty in calculated $M^{0\nu}$ of different groups, it is of great importance if there are experimental means allowing to measure these matrix elements. In a recent work Ref.~\cite{Rod09} a proposal is put forward which shows that such a measurement of the Fermi part $M_{F}^{0\nu}$ of the total matrix element $M^{0\nu}$ is in principle possible. Here, we would like to briefly discuss this result.

A similarity between the $0\nu\beta\beta$ neutrino potential and the radial dependence of the two-body Coulomb interaction is exploited in Ref.~\cite{Rod09}. When in addition one makes use of the isospin 
conservation by strong interaction, the matrix element $M_{F}^{0\nu}$ can be transformed as
to acquire the form of an energy-weighted double Fermi transition matrix element which is dominated
by the amplitude of the double Fermi transition via the IAS in the intermediate nucleus into the ground state of the final nucleus:
\begin{equation}
M^{0\nu}_F \approx - \frac{2}{e^2}\,\bar\omega_{IAS} 
\langle 0_f | \hat T^{-} |IAS \rangle  \langle IAS | \hat T^{-} |0_i\rangle .
\label{MFappr}
\end{equation}
Here, the second Fermi transition amplitude is due to an admixture of the double IAS 
in the final nucleus to the ground state of the parent nucleus: 
$\langle 0_f |\hat T^{-}| IAS\rangle \langle IAS |\hat T^{-}|0_i\rangle=\langle 0_f | DIAS\rangle \langle DIAS |\hat T^{-}| IAS\rangle \langle IAS |\hat T^{-}|0_i\rangle$. 

Therefore, the total $M^{0\nu}_F$ can be reconstructed according to Eq.~(\ref{MFappr}), 
if one is able to measure the $\Delta T=2$ isospin-forbidden m.e. 
$\langle 0_f | \hat T^{-} | IAS \rangle$, for instance in charge-exchange reactions of the $(n,p)$-type.
Using recent QRPA calculation results for $M^{0\nu}_F$, this m.e. can roughly be estimated as 
$\langle 0_f | \hat T^{-} | IAS \rangle \sim 0.005$,  
i.e. about a thousand times smaller than the first-leg m.e. $\langle IAS | \hat T^{-} | 0_i \rangle \approx \sqrt{N-Z}$. 
This strong suppression of $\langle 0_f | \hat T^{-} | IAS \rangle$ reflects smallness of the isospin mixing effects in nuclei. 
The IAS has been observed as a prominent and extremely narrow resonance and its various features have well been studied 
by means of (p,n), ($^3$He,t) and other charge-exchange reactions.
This gives us a hope that a measurement of $\langle 0_f | \hat T^{-} | IAS \rangle$ in the (n,p) charge-exchange channel might be possible. More generally, a measurement by whichever experimental mean of the $\Delta T=2$ admixture of the DIAS in the final ground state would be enough to determine $M_{F}^{0\nu}$.

Of course, by measuring only $M_{F}^{0\nu}$ one does not get the total m.e. $M^{0\nu}$ but rather its 
sub-leading contribution. However, knowledge of $M_{F}^{0\nu}$ itself brings a very important piece of information, since it can help to discriminate between  
different nuclear structure models in which calculated $M_{F}^{0\nu}$ may differ by as much as a factor of 5.
In addition, the ratio $M_{F}^{0\nu}/M_{GT}^{0\nu}$ may be more 
reliably calculable in different models than $M_{F}^{0\nu}$ and $M_{GT}^{0\nu}$ separately. Let us put forward here some simple arguments in support of the latter statement. 
Since only small internucleon distances determine $M^{0\nu}$, 
then only nucleon pairs in the spatial relative $s$-wave must dominantly contribute to the m.e.. 
The isotensor Coulomb interaction only couples $T=1$ pairs which must then be in the state with the total spin $S=0$ to assure antisymmetry of the total two-body wave function. 
Because of this and the fact that $\mbox{\boldmath $\sigma$}_1 \cdot \mbox{\boldmath $\sigma$}_2|S=0,T=1 \rangle = -3 |S=0,T=1 \rangle$, a natural estimate for the Gamow-Teller m.e. is 
$M^{0\nu}_{GT}=-3 M^{0\nu}_F$ 
provided the neutrino potential is the same in both F and GT cases.
The high-order terms of the nucleon weak current which are present in the case of the GT m.e., but absent in the F m.e.,
change a bit this simple estimate to $M_{GT}^{0\nu}/M_{F}^{0\nu}\approx -2.5$. Also, an uncertainty of few per cents 
may come from the difference in the mean nuclear excitation energies in the F and GT cases. 
It is worth noting that the recent QRPA results~\cite{Rod03a,Rod08,Kort07} 
are in good correspondence with these simple estimates.

\section{Conclusion}
In this contribution the present status of calculations of the nuclear matrix elements $M^{0\nu}$ for neutrinoless double beta decay is reviewed. The matrix elements $M^{0\nu}$ of different groups calculated within 
the QRPA seem to converge. At the same time, the $M^{0\nu}$ of the SM are substantially smaller (up to factor 2 for lighter nuclei like $^{76}$Ge) and this discrepancy is under active discussion in literature now. 
The recent results obtained within the IBM method agree surprisingly well with the QRPA ones.
In addition, we have briefly described a proposal which allows in principle to measure the neutrinoless double beta decay Fermi matrix element.

{\bf Acknowledgement:} Many of the original results reported here have been obtained in a long-time fruitful collaboration with Amand Faessler,   Fedor \v{S}imkovic and Petr Vogel, which is gratefully acknowledged by the author. 
The work is supported in part by the DFG
within the SFB TR27 ``Neutrinos and Beyond''.

\end{document}